\documentclass{article}
\pdfoutput=1
\usepackage{graphicx}  
\usepackage{amsmath}   
\usepackage[compress]{cite}
\usepackage{amssymb}   
\usepackage{bm} 
\usepackage{dcolumn}
\usepackage{color}
\usepackage[caption=false]{subfig}
\usepackage{mathrsfs}
\usepackage{amsfonts}
\usepackage{varioref}
\usepackage{float}
\restylefloat{table}
\usepackage[caption=false]{subfig}
\RequirePackage[colorlinks,citecolor=blue,urlcolor=magenta,linkcolor=blue]{hyperref}
\input epsf
\usepackage[labelsep=none,labelformat=empty]{caption}
\def\gr{general relativity}

\labelformat{section}{Section #1} 
\labelformat{subsection}{Section #1} 
\labelformat{subsubsection}{Section #1}
\labelformat{subsubsubsection}{Section #1}
\labelformat{equation}{Eq.~(#1)} 
\labelformat{figure}{Fig.~#1} 
\labelformat{subfigure}{Fig.~\thefigure#1} 
\labelformat{table}{Table.~#1} 
\labelformat{appendix}{Appendix #1}

\title{Signatures of regular black holes from the quasar continuum spectrum}
\author{Indrani Banerjee\footnote{banerjeein@nitrkl.ac.in} ~$^{1}$\\
{\small{$^{1}$Department of Physics and Astronomy, National Institute of Technology, Rourkela-769008, India}}
}

\begin{document}

\maketitle

\begin{abstract}
Regular black holes arising in Einstein gravity coupled to non-linear electrodynamics are worth studying as they can circumvent the $r=0$ curvature singularity arising in general relativity. In this work we explore the signatures of regular black holes with a Minkowski core from the quasar continuum spectrum. We use thin-disk approximation to derive the theoretical luminosity from the accretion disk and compare it with the optical data of eighty Palomar Green quasars. Our analysis based on error estimators like the $\chi^2$, the Nash-Sutcliffe efficiency, the index of agreement etc. reveal that optical observations of quasars favor the Kerr scenario compared to black holes in non-linear electrodynamics. The implications are discussed.

\end{abstract}

\section{Introduction}\label{S1}
The success of general relativity (GR) in explaining observations at all length scales makes it the most widely accepted theory of gravity till date \cite{Will:2005yc,Will:1993ns,Will:2005va}. With two successive ground breaking discoveries, namely, the gravitational waves from merging compact binaries \cite{Abbott:2017vtc,TheLIGOScientific:2016pea,Abbott:2016nmj,TheLIGOScientific:2016src,Abbott:2016blz} and the image of black holes \cite{Fish:2016jil,Akiyama:2019cqa,Akiyama:2019brx,Akiyama:2019sww,Akiyama:2019bqs,Akiyama:2019fyp,Akiyama:2019eap}, the theory has received observational confirmation in the strong field regime as well. Despite its astounding success, the theory has certain shortcomings, e.g it cannot explain the flat rotation curves of galaxies and the accelerated expansion of the universe \cite{Milgrom:1983pn,Milgrom:2003ui,Bekenstein:1984tv,Perlmutter:1998np,Riess:1998cb} and is marred with singularities \cite{Hawking:1973uf} associated with the big bang and black holes. This has given birth to a large number of alternatives to GR, which involves modification in the gravity sector, addition of extra fields in the matter sector or both \cite{Nojiri:2003ft,Nojiri:2006gh,Capozziello:2006dj,Lanczos:1932zz,Lanczos:1938sf,Lovelock:1971yv,Padmanabhan:2013xyr,Shiromizu:1999wj,Dadhich:2000am,Harko:2004ui,Carames:2012gr,Horndeski:1974wa,Sotiriou:2013qea,Babichev:2016rlq,Charmousis:2015txa}. 

The celebrated theorems of Hawking and Penrose \cite{Hawking:1973uf} state that spacetime singularities are inevitable in GR. It is believed that suitable quantum gravity models can evade the $r=0$ curvature singularity present in general relativistic black holes. However, the quantum nature of gravity continues to be an ill-understood subject till date \cite{Horava:1995qa,Horava:1996ma,Polchinski:1998rq,Polchinski:1998rr,Ashtekar:2006rx,Ashtekar:2006uz,Ashtekar:2006wn,Kothawala:2013maa,Kothawala:2015qxa}. In the absence of a well accepted theory of quantum gravity, the black hole singularity issue can be addressed classically by invoking regular black holes \cite{Ansoldi:2008jw,1966JETP...22..378G,Fan:2016hvf}. One possible scenario where such black hole solutions arise comprise of Einstein gravity coupled to non-linear electrodynamics. Such black holes have a horizon surrounding $r=0$ and the curvature invariants are finite at all points in spacetime \cite{Bronnikov:2000vy,Borde:1996df,Barrabes:1995nk,Ayon-Beato:1999qin,Bonanno:2000ep,Nicolini:2005vd,Myung:2007qt,PhysRevLett.96.031103}. The core at $r=0$ often assumes a de Sitter character \cite{Frolov:1988vj,Mukhanov:1991zn,Brandenberger:1993ef} which has been studied quite extensively. It is believed that such a black hole may be formed when unlimited increase of spacetime curvature during a stellar collapse gets halted due to quantum fluctuations. 

Apart from a de Sitter nature, the core at $r=0$ may also be of Minkowski type which arises as solution of Einstein's equations with an anisotropic fluid as the source. Such a fluid obeys the weak energy condition and resembles the Maxwell's stress tensor far from the black hole \cite{Culetu:2014lca}. 
The electric field therefore resembles the Coulomb field and the static, spherically symmetric metric assumes the form of the Reissner Nordstr\"{o}m spacetime at large distances from the source. Investigating the observational characteristics of such a spacetime is important as the mass of the black hole has an exponential convergence factor which makes the quantum gravity model associated with the classical theory finite to all orders \cite{Brown:1980uk}. Studying a finite quantum gravity theory is important as it can not only address the singularity issues in GR, but can also resolve the cosmological constant problem \cite{Moffat:2001jf} and eradicate the divergences arising in flat space quantum field theories.

Since we are interested in exploring the observational properties of the regular black hole with a Minkowski core, we consider the stationary, axisymmetric counterpart of the aforesaid spacetime. The rotating solution is obtained by applying the Newman-Janis algorithm \cite{Newman:1965tw} or other mathematical techniques \cite{Azreg-Ainou:2014pra,Azreg-Ainou:2014aqa,Azreg-Ainou:2014nra} to the static, spherically symmetric seed metric. This has been worked out by Ghosh \cite{Ghosh:2014pba} which we use in this work. 
Needless to say, such a spacetime has a horizon covering the $r=0$ Minkowski core and resembles the Kerr-Newman metric at large distances from the source \cite{Ghosh:2014pba}.
 
The goal of the present work is to decipher the signatures of the aforesaid spacetime from the continuum spectrum of a sample of eighty Palomar Green quasars \cite{Davis:2010uq}. We assume that these quasars are regular black holes with a Minkowski core arising in Einstein gravity coupled to non-linear electrodynamics. Considering the rotating solution \cite{Ghosh:2014pba} we derive the luminosity from the accretion disk of these quasars in the thin-disk approximation \cite{Novikov_Thorne_1973,Page:1974he} and compare it with their observed optical luminosities. This in turn enables us to establish constrains on the non-linear electrodynamics charge parameter supposed to be associated with the quasars. Our analysis 
is based on error estimators like the chi-square, the Nash-Sutcliffe efficiency, the index of agreement and their modified forms.
We also provide independent estimates of spin for some of the quasars. 

We organise the paper in the following way: \ref{S2} is dedicated in describing the regular black holes with a Minkowski core. In \ref{S3} we derive the theoretical luminosity from the accretion disk in the aforesaid background using thin-disk approximation. The theoretical optical luminosity is then compared with the observed ones 
using the error estimators in \ref{S4}. We conclude with a summary of our results with some scope for future work in \ref{S5}. Here we assume (-,+,+,+) as metric convention and work with geometrized units which considers G=c=1.

\section{Regular black holes with a Minkowski core}
\label{S2}
In this section we describe regular black holes with an asymptotically Minkowski core which arise in Einstein gravity coupled to non-linear electrodynamics. In this case, Einstein's equations is sourced by an anisotropic fluid whose energy-momentum tensor assume the form: 
\begin{align}
&T^0_0=-\rho(r)=\frac{-\mathcal{M} k}{4\pi  r^4} e^{-k/{r}};\nonumber \\
&T^1_1=-\rho(r)=\frac{-\mathcal{M} k}{4\pi  r^4} e^{-k/{r}}; \nonumber \\
&T^2_2=T^3_3=\frac{\mathcal{M} k}{4\pi  r^4}\bigg(1-\frac{k}{2r}\bigg)e^{-k/r}
\label{energy}
\end{align}
where $k=Q^2/2\mathcal{M}$ such that $Q$ is the charge and $\mathcal{M}$ is the mass of the black hole. It is interesting to note that the energy momentum tensor in \ref{energy}
is consistent with the weak energy condition and resembles the Maxwell stress tensor far from the source \cite{Culetu:2013fsa}. 

With the source being given by \ref{energy} the static, spherically symmetric and asymptotically flat, black hole solution of Einstein's equations assume the form,
\begin{align}
ds^2=-(1-\frac{2\mathcal{M}}{r}e^{-k/r}) dt^2 + \frac{dr^2}{1-\frac{2\mathcal{M}}{r}e^{-k/r}} + r^2 d\theta^2 + r^2sin^2 \theta d\phi^2
\label{metric}
\end{align}
Since we are interested in studying astrophysics of these black holes, we need to consider the rotating counterpart of \ref{metric}.
This is obtained by applying the Newman-Janis algorithm to the above static, spherically symmetric seed metric \cite{Ghosh:2014pba}. This gives rise to the stationary, axisymmetric counterpart of \ref{metric} which in Boyer-Lindquist coordinates is given by,
\begin{align}
\label{2}
ds^{2} &=-\bigg{(} 1 - \frac{2{M}(r)r}{{\Sigma}}\bigg{)}dt^{2} - \frac{4{a}{M}(r)r}{{\Sigma}}\sin^{2}\theta dt d\phi + \frac{{\Sigma}}{\Delta}dr^{2} \nonumber\\
&+{\Sigma} d\theta^{2} + \bigg{(} r^{2} + {a}^{2} + \frac{2{M}(r)r{a}^{2}}{{\Sigma}}\sin^{2}\theta\bigg{)}\sin^{2}\theta d\phi^{2}
\end{align}
where, 
\begin{align}\label{3}
{\Sigma} = r^{2} + {a}^{2}\cos^{2}\theta ~ {,} ~ \Delta = r^{2} + {a}^{2} - 2{M}(r)r
\end{align}
and $a$ is the angular momentum of the black hole.
In \ref{2}, $M(r)$ is the mass function given by,  
\begin{align}
\label{4}
M(r) = \mathcal{M}e^{-k/r}
\end{align}
which becomes $\mathcal{M}$ as $r\to \infty$. $M(r)$ can be thought of as mass inside the sphere of radius $r$. It is important to note that the metrics in \ref{metric} and \ref{2} are regular at all points in the spacetime. At $r=0$, the energy density vanishes unlike black holes with a de Sitter core where the energy density becomes constant at $r=0$. Moreover, the metric in \ref{metric} or \ref{2} reduces to the Minkowski background at $r=0$. 
When $k=0$, the metric in \ref{metric} resembles the Schwarzschild spacetime while the metric in \ref{2} reduces to the Kerr spacetime. For a non-zero $k$ with large $r$, in the limit $r>>k$, ${M}(r)=\mathcal{M}-\frac{Q^2}{2r}$ and the metric in \ref{2} assumes the form of the Kerr-Newman spacetime. Studying the regular solution in \ref{metric} or \ref{2} is worthwhile as the curvature invariants become immensely simpler than black hole spacetimes with a regular de Sitter core, e.g. Bardeen metric, and bears several physically interesting features defined by the Lambert W function \cite{Valluri:2000zz,Boonserm:2008zg,Boonserm:2010px,Boonserm:2013dua,Boonserm:2018orb,Sonoda:2013kia,Sonoda:2013jia,Culetu:2013fsa}.

The event horizon of the spacetime given by \ref{2} has a spherical topology and is obtained by solving for $r$ in the equation,
\begin{align}
\label{5}
g^{rr}=\Delta=r^2 + a^2 - 2r\mathcal{M}e^{-k/r}=0
\end{align}
The existence of an event horizon requires one to solve for real, positive values of $r$ in \ref{5}. The requirement of an event horizon or a black hole solution imposes restrictions on the values of $k$ and $a$, \cite{Kumar:2018ple} such that $0\lesssim k\lesssim 0.7$ and for every $k$, the spin $a$ lies in a certain range.
In the present work we shall constrain both $k$ and $a$ from the quasar continuum spectrum.

\section{Continuum spectrum from the accretion disk in a stationary, axisymmetric spacetime}
\label{S3}
The continuum spectrum emitted by the accretion disk surrounding a black hole is sensitive to the background spacetime and the properties of the accretion flow. Therefore it serves as an important astrophysical tool to decipher the nature of the background metric.
Consider the stationary, axisymmetric spacetime which is asymptotically flat and has reflection symmetry,
\begin{align}
ds^2=g_{tt}dt^2 + 2g_{t\phi}dt d\phi + g_{\phi\phi}d\phi ^2 + g_{rr}dr^2 +g_{\theta\theta} d\theta^2~,
\label{3-1}
\end{align}
For such a spacetime the metric components are independent of $t$ and $\phi$, such that we have two conserved quantities, namely specific energy $\mathcal{E}$ and specific angular momentum $\mathcal{L}$ which for timelike particles are respectively given by,
\begin{align}
\label{3-2}
{\mathcal{E}} &= \frac{-g_{tt} - {\Omega}g_{t\phi}}{\sqrt{-g_{tt} - 2{\Omega}g_{t\phi} - {\Omega}^2 g_{\phi\phi}}} \\
{\mathcal{L}} &= \frac{{\Omega}g_{\phi\phi} + g_{t\phi}}{\sqrt{-g_{tt} - 2{\Omega} g_{t\phi} - {\Omega}^{2} g_{\phi\phi}}}
\end{align}
where, $\Omega$ is the angular velocity of massive test particles,
\begin{align}
\label{3-3}
{\Omega} = \frac{d\phi}{dt} = \frac{-g_{t\phi,r} \pm \sqrt{(-g_{t\phi,r})^2 - (g_{\phi\phi,r})(g_{tt,r})}}{g_{\phi\phi,r}}
\end{align}
It is important to note that we consider motion along the equatorial plane such that ${\mathcal{E}}$ and ${\mathcal{L}}$ are functions of $r$ and the conserved quantities are independent of $g_{\theta \theta}$.

In this section we shall derive the continuum spectrum from the accretion disk surrounding a rotating black hole with line element given by \ref{3-1}. While deriving the spectrum we shall consider the thin-disk approximation  \cite{Novikov_Thorne_1973,Page:1974he} where the accretion flow is confined along the 
equatorial plane ($\theta = \pi/2$). As matter inspirals and falls into the central black hole the azimuthal velocity $v_{\phi}$ far exceeds the radial velocity $v_{r}$ and the vertical velocity $v_{z}$ such that $v_{\phi} \gg v_{r} \gg v_{z}$. Because of this condition, the system does not harbor outflows and motion is roughly along circular geodesics. The presence of viscosity in the system endows minimal radial velocity to the accreting fluid, such that the matter slowly inspirals and falls into the black hole.   
Due to circular geodesic motion along the equatorial plane the disk height $h(r)\ll r$ where $r$ is the radial distance from the central black hole. The energy-momentum tensor of the accreting fluid assumes the form,
\begin{align}
\label{3-4}
\mathcal{T}^{\mu}_{\nu} = \rho(1 + {\Pi})u^{\mu}u_{\nu} + s^{\mu}_{\nu} + u^{\mu}v_{\nu} + v^{\mu}u_{\nu}
\end{align}
where $\rho$ is the proper density and $u_{\nu}$ the four velocity of the accreting fluid. In \ref{3-4} ${\Pi}$ represents the specific internal energy of the accreting fluid such $\rho{\Pi}u^{\mu}u_{\nu}$ is the dissipation term. The thin-disk approximation implies that ${\Pi}\ll 1$ which in turn ascertains that the rest energy of the accreting fluid far supercedes the dissipation effects. As a consequence, special relativistic effects due to local thermodynamic, hydrodynamic and radiative properties of the flow can be neglected and no heat is retained with the flow. However, the general relativistic effects due to the black hole continues to play a conspicuous role. $s^{\mu\nu}$ is associated with the the stress-tensor while $v^{\mu}$ represents the energy flux relative to the local inertial frame. Note that $s^{\mu\nu}$ and $v^{\mu}$ are orthogonal to the 4-velocity, i.e., $s_{\mu \nu}u^{\mu}=0=v_{\mu}u^{\mu}$. 
The photons generated due to viscous dissipation interacts very efficiently with the accreting fluid before leaving the system as $v^z$. Thus, every annulus of the disk emits a black body radiation such that the continuum spectrum from the accretion disk in the thin-disk approximation represents a multi-color black body spectrum.

We now aim to calculate the flux and subsequently the luminosity from the accretion disk in the thin-disk approximation. We shall assume that accretion takes place at a steady rate $\dot{M}$ which is very small compared to the mass of the black hole such that the process of accretion does not change the black hole mass over the timescales of interest.
The metric therefore continues to remain stationary and axisymmetric and is given by \ref{2}. The equation governing conservation of mass is given by,
\begin{align}
\label{3-5}
\dot{M} = -2\pi\sqrt{-\gamma}u^{r}{\Sigma}
\end{align}
where $u_r$ is the radial velocity, $\Sigma=\int \rho dz$ is the average surface density of the accreting fluid and $\gamma$ is the determinant of the two dimensional metric associated with the equatorial plane such that $\gamma = -g_{rr}(g^{2}_{t\phi} - g_{tt}g_{\phi\phi})$. The accretion flow also conserves angular momentum and energy. These are best described by defining a current $J^\nu_i$ where $i\equiv t,\phi$ corresponding to energy and angular momentum respectively. The conservation of energy thus assumes the form 
\begin{align}
\label{3-6}
\nabla_\nu J^\nu_t=\nabla_\nu \bigg \lbrace T^\nu_\mu \bigg(\frac{\partial}{\partial t}\bigg)^\mu\bigg\rbrace=
\partial_{r}(\dot{M}{\mathcal{E}} - 2\pi\sqrt{-\gamma}{\Omega} w^{r}_{\phi}) ~- 4\pi\sqrt{-\gamma}\mathcal{F}{\mathcal{E}}=0
\end{align}
while the conservation of angular momentum is governed by,
\begin{align}
\label{3-7}
\nabla_\nu J^\nu_\phi=\nabla_\nu \bigg\lbrace T^\nu_\mu \bigg(\frac{\partial}{\partial \phi}\bigg)^\mu\bigg\rbrace=\partial_{r}(\dot{M}{\mathcal{L}} - 2\pi\sqrt{-\gamma}w^{r}_{\phi})~- 4\pi\sqrt{-\gamma}\mathcal{F}{\mathcal{L}}=0
\end{align}
In \ref{3-6} and \ref{3-7} $w^{r}_{\phi}$ refers to the height and time averaged stress tensor in the local rest frame of the accreting fluid,
\begin{align}\label{3-8}
w^{\alpha}_{\beta} = \int^{h}_{-h} dz \langle s^{\alpha}_{\beta} \rangle
\end{align}
while $\mathcal{F}$ represents the flux radiated from the accretion disk,
\begin{align}\label{3-9}
\mathcal{F} \equiv \langle v^{z}(r,h)\rangle = \langle -v^{z}(r,-h)\rangle
\end{align}
For a detailed derivation of the conservation laws one is referred to \cite{Novikov_Thorne_1973,Page:1974he}. 
From the conservation laws one can derive an analytic expression for the flux \cite{Page:1974he} which assumes the form,
\begin{align}
\mathcal{F} &= \frac{\dot{M}}{4\pi\sqrt{-\gamma}}f \label{3-10}~~~~\rm {where} \\
f &=~ - \frac{{\Omega}_{,r}}{({\mathcal{E}} - {\Omega}{\mathcal{L}})^2} \bigg{[} {\mathcal{E}}{\mathcal{L}} - {\mathcal{E}}_{ms} {\mathcal{L}}_{ms} - 2\int^{r}_{r_{ms}} {\mathcal{L}}{\mathcal{E}}_{,r^{\prime}} dr^{\prime} \bigg{]} \label{3-11}
\end{align}
In \ref{3-11}, $\mathcal{E}_{ms}$ and $\mathcal{L}_{ms}$ correspond to the specific energy and specific angular momentum at the marginally stable circular orbit $r_{ms}$. It is important to note that while deriving $\mathcal{F}$ we assume that the viscous stress $w^{r}_{\phi}$ vanishes at $r_{ms}$ such that the azimuthal velocity drops to zero and the accreting matter plunges into the black hole with pure radial velocity.
In order to obtain the radius of the marginally stable circular orbit we need to understand the nature of the effective potential for motion of massive test particles. Such a potential is obtained from the equation $p^\mu p_\mu=-m^2$ and is given by,
\begin{align}\label{3-12}
V_{eff}(r) ~ = ~ \frac{{\mathcal{E}}^2 g_{\phi\phi} + 2{\mathcal{E}} {\mathcal{L}} g_{t\phi} + {\mathcal{L}}^2g_{tt}}{g_{t\phi}^2 - g_{tt}g_{\phi\phi}} ~ - ~ 1
\end{align}
Such a potential has an inflection point and the radius of the marginally stable circular orbit is obtained by solving for $V_{eff}=V^\prime_{eff}=V^{\prime\prime}_{eff}=0$ where prime denotes derivative with respect to $r$. 
The photons emitted in the disk due to viscous dissipation undergo repeated collisions with the accreting matter before leaving the system. Consequently each annulus of the disk emits a black body spectrum, such that the thin accretion disk is `geometrically thin but optically thick'. The total emission from the disk is an envelope of the individual black body spectrum and is known as the multicolor black body spectrum. The temperature profile at any given radius therefore satisfies the Stefan-Boltzmann's law, $T(r)=\lbrace F(r)/\sigma\rbrace^{1/4}$ where $\sigma$ is the Stefan-Boltzmann constant, and ${F}(r)=\mathcal{F}(r)c^6/G^2\mathcal{M}^2$. Therefore at every radius the disk emits a Planck spectrum, such that the luminosity from the disk at an observed frequency $\nu$ is given by,
\begin{align}\label{3-13}
L_{\nu} = 8\pi^2 r_{g}^2 \cos{i} \int^{r_{out}}_{r_{ms}} \sqrt{g_{rr}} B_{\nu}(T(r))r~ dr
\end{align}
While evaluating $L_\nu$ we make our radial coordinate dimensionless such that $r\equiv r/r_g$ where $r_g$ is the gravitational radius given by $r_g = GM/c^2$. In \ref{3-13}
$i$ is the inclination angle while the Plank spectrum is given by,
\begin{align}\label{3-14}
B_{\nu} (T(r)) = \frac{2h\nu^3}{c^2\big{[} exp(\frac{h\nu} { z_{g}kT(r)}) -1\big{]}}
\end{align}
In \ref{3-14},the gravitational redshift factor $z_g$ associated with the outgoing photons is given by,
\begin{align}\label{3-15}
z_{\rm g}=E\frac{\sqrt{-g_{tt}-2{\Omega} g_{t\phi}-{\Omega}^2 g_{\phi\phi}}}{{E}-\Omega {L}} 
\end{align}
which takes into account the change in frequency suffered by the photon while travelling from the emitting medium to the observer \cite{Ayzenberg:2017ufk}. In \ref{3-15}, $E$ and $L$ correspond to the specific energy and specific angular momentum of the outgoing photon.  

One may note from \ref{3-13} that $L_{\nu}$ is sensitive to the mass of the black hole, the accretion rate, the inclination angle and the metric components which are given by \ref{2}.

\begin{figure}[t!]
\begin{center}
\hspace{-2.1cm}
\includegraphics[scale=0.69]{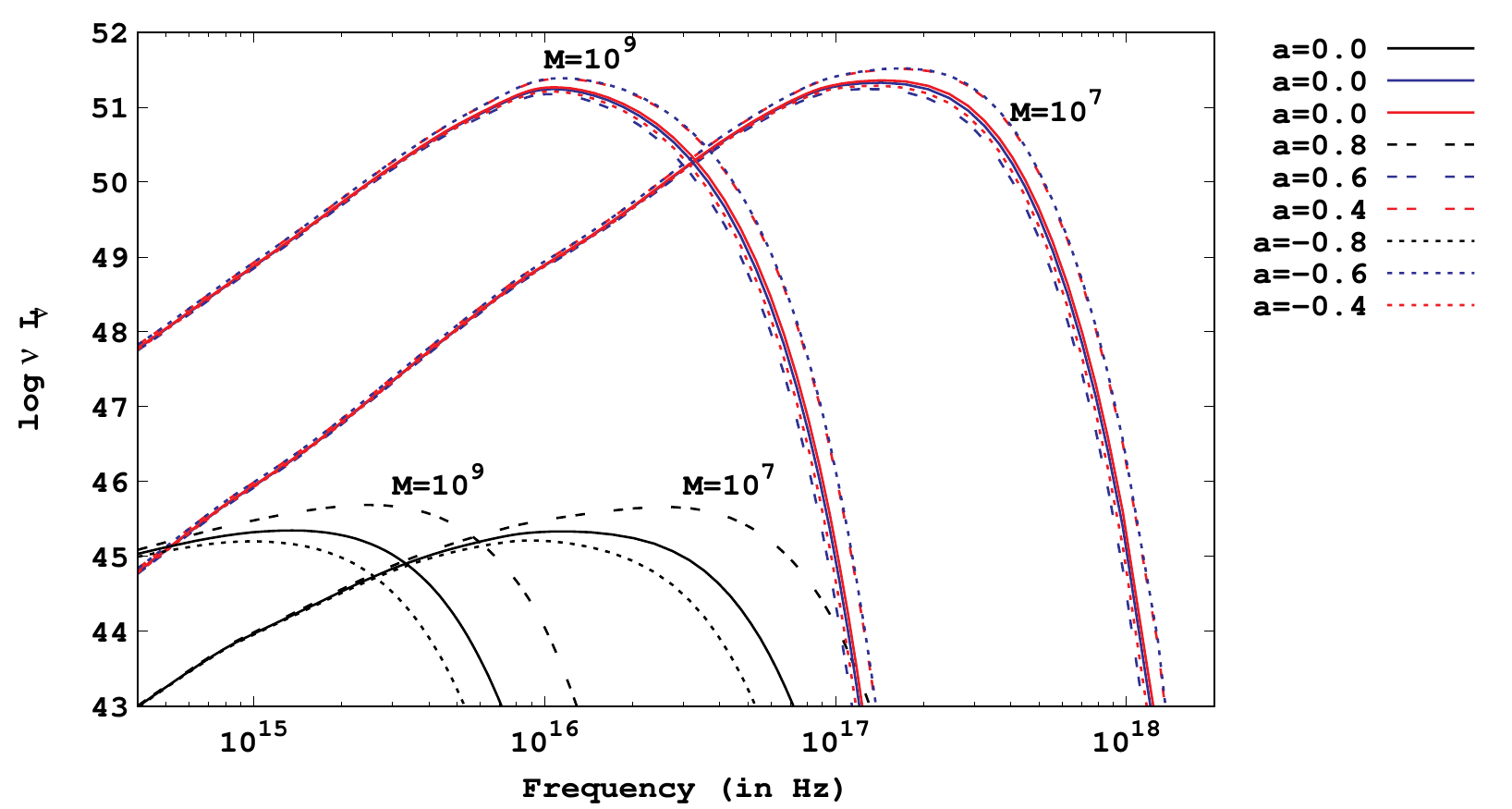}
\end{center}
\caption{Figure 1: The figure above depicts the variation of the theoretical luminosity from the accretion disk with frequency considering two sets of black hole masses, namely, $M=10^9 M_\odot$ and $M=10^7 M_\odot$. The inclination angle taken to be $ i=cos^{-1}0.8$ while the accretion rate assumed is $1 M_\odot~ \rm year^{-1}$.
The black, blue and red lines correspond to $k=0$, $k=0.3$ and $k=0.5$, where $k$ is the non-linear electrodynamics charge parameter.
Keeping $k$ fixed, non-rotating black holes are represented by solid lines, while dashed and dotted lines correspond to prograde and retrograde black holes.
}
\label{Fig1}
\end{figure}
In \ref{Fig1} we plot the dependence of the luminosity from the accretion disk on frequency and the metric parameters $k$ and $a$. We consider two sets of black holes masses, e.g. $\mathcal{M}=10^7 M_\odot$ and $\mathcal{M}=10^9 M_\odot$. For both the masses the accretion rate is taken to be $1 M_\odot~ \rm year^{-1}$ while the inclination angle is assumed to be $i=\cos^{-1}0.8$. We note from the figure that for any given mass the effect of the metric becomes conspicuous in the high frequency regime (spectra becomes distinguishable). This is because the high frequency part of the spectrum is emitted by the hot inner disk where the effect of the background metric is generally expected to be predominant. We further note from \ref{Fig1} that the accretion disk surrounding a lower mass black hole emits higher frequencies. This is because the temperature of the disk scales as $T(r)\propto \mathcal{M}^{-1/4}$. Therefore, a lower mass black hole has a hotter accretion disk and hence emits higher frequencies. In \ref{Fig1} the spectra denoted by black, blue and red lines correspond to $k=0$, $k=0.3$ and $k=0.5$. In each case $k$ and $a$ are chosen in such a way that the event horizon remains real and positive \cite{Kumar:2018ple}. The prograde spins are denoted by dashed lines, the retrograde spins are denoted by dotted lines while the non-rotating black holes are denoted by solid lines. From the figure we note that accretion disk around black holes with a non-zero non-linear electrodynamics charge are much more luminous than Kerr black holes. This will play a crucial role in determining the observationally favored value of $k$.

\section{Contact with observations }
\label{S4}
 
In this section we compute the optical luminosity at 4861\AA\ of a sample of eighty Palomar Green quasars studied in Davis \& Laor \cite{Davis:2010uq} using the thin disk approximation. In order to compute the theoretical optical luminosity we need to provide information regarding the mass, inclination angle and the accretion rate (\ref{3-13}). The masses of these quasars have been determined by reverberation mapping \cite{Kaspi:1999pz,Kaspi:2005wx,Boroson:1992cf,Peterson:2004nu} and are reported in Davis \& Laor \cite{Davis:2010uq}. For some quasars masses based on $M-\sigma$ method are also mentioned \cite{Davis:2010uq}. Here we shall consider masses determined by the method of reverberation mapping. Using the observed spectra in the  optical \cite{1987ApJS...63..615N}, UV \cite{Baskin:2004wn}, far-UV \cite{Scott:2004sv}, and soft X-ray \cite{Brandt:1999cm} domain the bolometric luminosity of these quasars are estimated \cite{Davis:2010uq}. Supermassive black holes are multi-component systems comprising of the disk, the corona, the jet and the dusty torus \cite{Brenneman:2013oba}. The accretion disk of these systems generally emit in the optical/UV part of the spectrum. Since emission from the UV regime has maximum contribution to the error in the bolometric luminosity we shall consider the observed optical luminosity reported in Davis \& Laor \cite{Davis:2010uq} for comparison with our theoretical estimates. 

Apart from mass, information about the accretion rate and the inclination angle are required to obtain the theoretical optical luminosity. The accretion rates of these quasars are reported in \cite{Davis:2010uq}. In order to determine the accretion rate several theoretical models have been considered \cite{Davis:2010uq}. The base model turns out to be the so called TLUSTY model which uses stellar-atmosphere-like assumptions of the disk structure with spin $a = 0.9$. However, they also provide the degree of variation of the accretion rate if other disk models or black hole spins are used. For example, TLUSTY model with spin $a = 0$ results in $40\%$ increment of the accretion rate for higher mass quasars 
and $10\%$ increase in the accretion rate for low mass quasars compared to the base model. When black body model with $a = 0.9$ is used the accretion rates turn out to be lower by $10\% - 20\%$ for all quasars. Similarly when black body model with $a = 0$ is used the accretion rate of low mass quasars decrease by 20\% while that of high mass quasars get enhanced by $40\%$ compared to the base model. Therefore irrespective of the choice of model or black hole spin we vary the accretion rate between $80\%$ to $140\%$ of the value reported in \cite{Davis:2010uq} for all the PG quasars. 

The quasar sample considered here are nearly face-on systems and therefore the inclination angle $i$ is varied between $\cos i \in \left(0.5,1\right)$ \cite{Antonucci:1993sg,Davis:2010uq,Wu:2013zqa}. This is consistent with the estimates of \cite{2017Ap&SS.362..231P} which reports the inclination angle for some quasars in the sample we are studying. 

Having discussed the inputs we use for the calculation of theoretical optical luminosity (i.e., mass, accretion rate and inclination angle), we now describe the procedure we adopt to constrain the non-linear electrodynamics charge parameter $k$:
\begin{itemize}
\item We consider the eighty quasars to be regular black holes with a Minkowski core. Therefore these objects should possess an event horizon. This is ensured by solving for $\Delta=0$ in \ref{3}. This gives us a theoretical restriction on the values of $k$ and $a$ such that $0\lesssim k\lesssim 0.7$ and for every $k$, $a$ lies in a certain range.
\item First of all, we choose a value of $k$ between $0\lesssim k\lesssim 0.7$. For the chosen $k$, the allowed range of $a$ gets fixed.
\item Next, we choose a quasar from the sample whose mass based on reverberation mapping is reported in \cite{Davis:2010uq}.
\item Keeping $k$ and $a$ fixed we now fix an inclination angle between  $0.5 \lesssim\cos i \lesssim 1$.
\item Keeping $k$, $a$ and $cos i$ fixed we vary the accretion rate between $0.8 \dot{M}$ to $1.4 \dot{M}$, where $\dot{M}$ is the accretion rate obtained from the base model and reported in \cite{Davis:2010uq}. 
\item For the chosen quasar and the given $k$ we compute the theoretical luminosity at 4861\AA\ for every combination of $a$, $\dot{M}$ and $\cos i$. The values of $a$, $\dot{M}$ and $\cos i$ which minimizes the error between the theoretical and the observed optical luminosities are noted and are referred to as $a_{min}$, $\dot{M}_{min}$ and $\cos i_{min}$ for the chosen quasar.
\item We repeat steps 3-6 for all the quasars and estimate the errors as discussed in the next section.
\item We then repeat steps 2-7 for all the values of $k$ in the physically allowed range.
\end{itemize} 
The present analysis assumes that all the eighty quasars have the same non-linear electrodynamics charge parameter $k$. Here $k$ essentially refers to the average charge and since $k$ is positive and varies in a very small range, this assumption is justified. We next consider several error estimators to arrive at the most favorable magnitude of $k$ derived from the optical observation of quasars.  

\subsection{Error estimators}
\label{S4-1}
\begin{itemize}
\item {\textbf {Chi-square} $\boldsymbol {\chi^{2}}~$}:~  
\begin{figure}[H]
\begin{center}
\hspace{-2.1cm}
\includegraphics[scale=0.67]{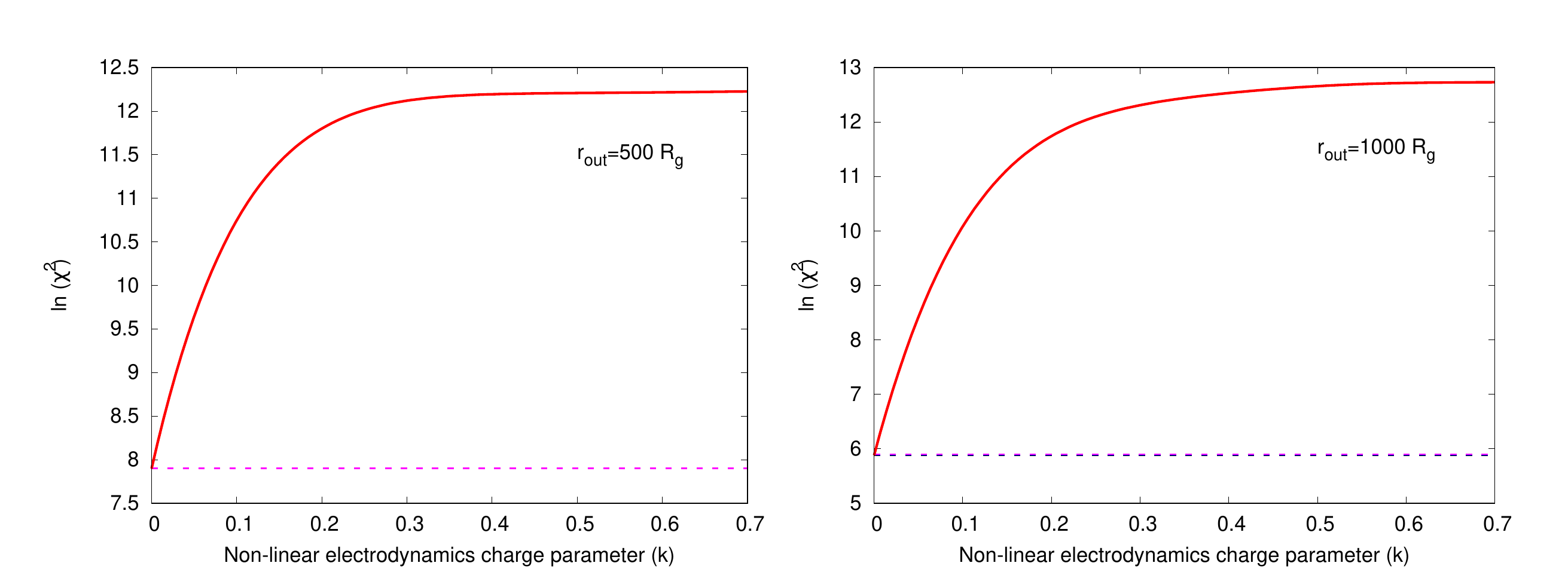}
\end{center}
\caption{Figure 2: The above figure illustrates the variation of $\chi^2$ with the non-linear electrodynamics charge parameter $k$. This is shown for two choices of $r_{out}$, namely, $r_{out}=500R_g$ (left panel) and $r_{out}=1000R_g$ (right panel). The figure clearly shows that $\chi^2$ minimizes for $k=0$ irrespective of the choice of $r_{out}$. We also note that for a non-zero $k$, $\chi^2$ rises very sharply such that the 1-$\sigma$, 2-$\sigma$ and 3-$\sigma$ contours (denoted by black, blue and magenta dashed lines) nearly overlap. This indicates that $k>0$ is ruled out outside 99\% confidence interval.
}
\label{Fig2}
\end{figure}
The $\chi^{2}$ of  a distribution is given by,
\begin{align}
\label{4-1}
\chi ^{2}\big(k)= \sum _{j=1}^{80}\frac{\left\{{O}_{j}-T_{j} \big(k,\lbrace a_{min,j}, \cos i_{min,j},\dot{M}_{min,j}\big\rbrace) \right\}^{2}}{\sigma _{j}^{2}}.
\end{align}
where $\{ {O}_{j}\}$ denotes the observed dataset and $T_j\big(k,\lbrace a_{min,j}, \cos i_{min,j},\dot{M}_{min,j}\rbrace\big)$ represents the theoretical optical luminosity of the $j^{th}$ quasar corresponding to $a_{min}$, $\cos i_{min}$ and $\dot{M}_{min}$ for a given $k$. The errors associated with the observed luminosity are denoted by $\sigma_j$. It is important to note that the errors corresponding to the observed optical luminosities of quasars are not given, therefore we use the errors associated with the bolometric luminosities \cite{Davis:2010uq} as the maximum possible error in the optical luminosity.

From \ref{4-1} it is easy to see that the value of $k$ minimizing the $\chi^2$ corresponds to the observationally favored magnitude of the non-linear electrodynamics charge parameter. The corresponding value of $\chi^2$ is denoted by $\chi^2_{min}$.
Further, it is important to note that we do not minimize the $\chi^2$ with respect to $k$, $a$, $\dot{M}$ and $cosi$ at the same time. This is because our primary goal is to determine the observationally favored value of the charge parameter $k$ from the quasar optical data. 
Thus, although $\chi^2$ depends on four parameters, the number of `interesting parameters' is one ($k$) and there are three `uninteresting parameters' ($a$, $\dot{M}$ and $cosi$) \cite{1976ApJ...210..642A}. When we have only one `interesting parameter' the 68\%, 90\% and 99\% confidence intervals are determined by $\Delta \chi^2=1, 2.71, 6.63$ from $\chi^2_{min}$. From \ref{3-13} we note that the theoretical luminosity depends both on the inner and the outer disk radius. The inner disk radius corresponds to $r_{ms}$ (which depends on $k$ and $a$) in the thin-disk approximation, whereas the outer disk radius is taken to be $r_{out}=500r_g$ and $r_{out}=1000r_g$.
The luminosity from the accretion disk is much more sensitive to the inner radius (when the metric is asymptotically flat) and therefore calculating the errors by varying $r_{out}$ is done just for completeness. The variation of $\chi^2$ with $k$ is shown in \ref{Fig2}. From the figure it is clear that $\chi^2$ minimizes for $k=0$ irrespective of the choice of $r_{out}$. Since the disk luminosity increases substantially for a non-zero $k$ (\ref{Fig1}), the $\chi^2$ increases sharply for a non-vanishing $k$. Therefore, the 1-$\sigma$, 2-$\sigma$ and 3-$\sigma$ contours denoted by black, blue and magenta lines almost overlap. Hence, non-vanishing $k$ are ruled out outside 3-$\sigma$ confidence interval. The $\chi^2$ analysis therefore tells us that optical observation of quasars favor the Kerr scenario compared to the regular black hole scenario with a Minkowski core.

\item \textbf{Nash-Sutcliffe Efficiency:}

The Nash-Sutcliffe Efficiency $E$ \cite{NASH1970282,WRCR:WRCR8013,2005AdG.....5...89K} is given by the ratio of the squared difference between the observed and the theoretical luminosity to the squared difference between the observed and average luminosity subtracted from unity. The mathematical expression corresponding to the Nash-Sutcliffe Efficiency $E$ is given by,
\begin{align}
\label{4-2}
E(k)=1-\frac{\sum _{j}\left\{{O}_{j}-T_{j}\bigg(k,\left \lbrace a_{min,j}, \cos i_{min,j},\dot{M}_{min,j} \right\rbrace \bigg)\right\}^{2}}{\sum _{j}\left\{{O}_{j}-{O}_{\rm av}\right\}^{2}}
\end{align}
\begin{figure}
\begin{center}
\hspace{-2.1cm}
\includegraphics[scale=0.67]{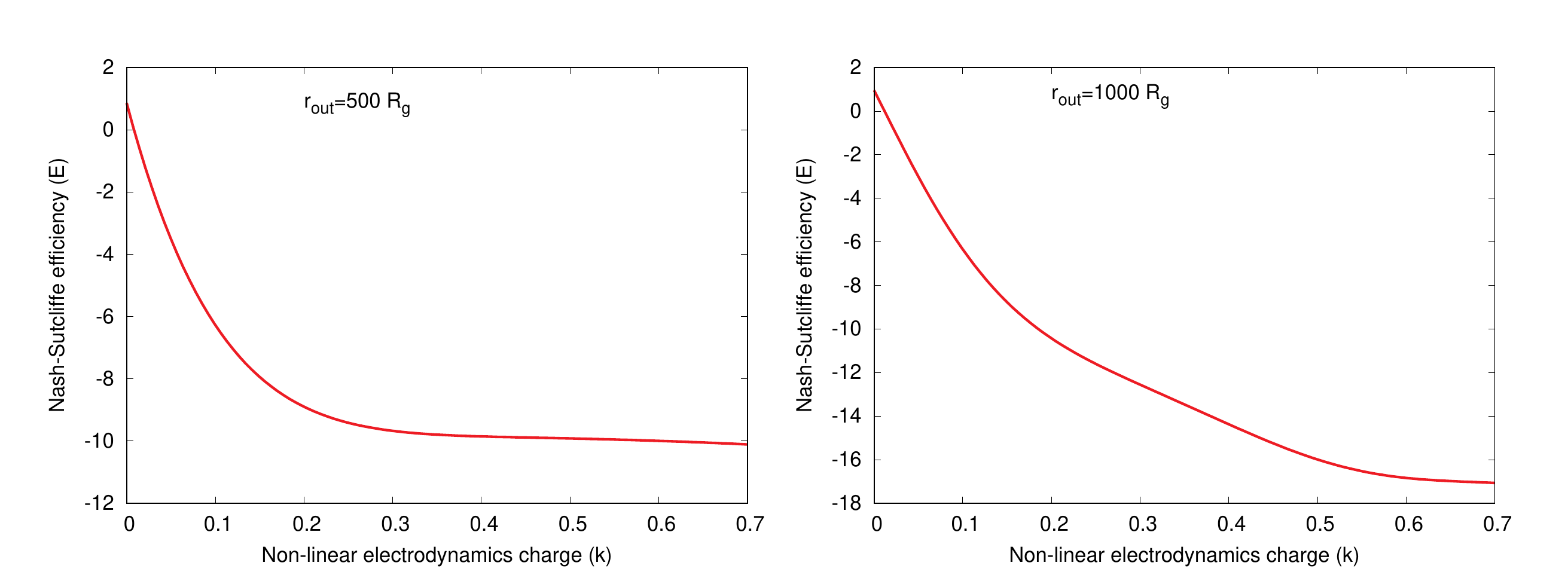}
\end{center}
\caption{Figure 3: The above figure illustrates the variation of the Nash-Sutcliffe Efficiency $E$ with the non-linear electrodynamics charge parameter $k$. In the left panel the theoretical optical luminosity is computed with $r_{out}=500r_g$ while the right panel corresponds to $r_{out}=1000r_g$ for computing the theoretical optical luminosity. The figure shows that $E$ maximizes for $k=0$ in agreement with our findings based on $\chi^2$ analysis.}
\label{Fig3}
\end{figure}

Here, ${O}_{\rm av}$ denotes the average value of the observed optical luminosities of the PG quasars. While computing the differences in \ref{4-2} we take logarithm of the observed, theoretical and average luminosity. From \ref{4-2} it is easy to note that $E$ can range from $-\infty ~\rm to ~ 1$. A negative $E$ implies that the mean of the observed data predicts the observation better than the theoretical model. 
From \ref{4-2} it is clear that the value of $k$ where $E$ maximizes corresponds to the observationally favored magnitude of $k$. \ref{Fig3} depicts the variation of $E$ with $k$ for two different choices of $r_{out}$. From the figure we note that $E$ maximizes when $k=0$ irrespective of the choice of $r_{out}$ which is in agreement with our results obtained from  $\chi^2$ estimate.

\item \textbf{Modified Nash-Sutcliffe Efficiency:} 
The presence of the squared term in the expression for Nash-Sutcliffe Efficiency makes it oversensitive to larger values of the luminosity. As a result, a modified version of the same (denoted by $E_1$) is put forward \cite{WRCR:WRCR8013} which is given by,
\begin{align}
\label{4-3}
E_{1}(k)&=1-\frac{\sum_{j}\big|{O}_{j}-T_{j}\big(k,\lbrace a_{min,j}, \cos i_{min,j},\dot{M}_{min,j} \rbrace \big)\big|}{\sum _{j}\big|{O}_{j}-{O}_{\rm av}\big|}
\end{align}

\begin{figure}
\begin{center}
\hspace{-2.1cm}
\includegraphics[scale=0.67]{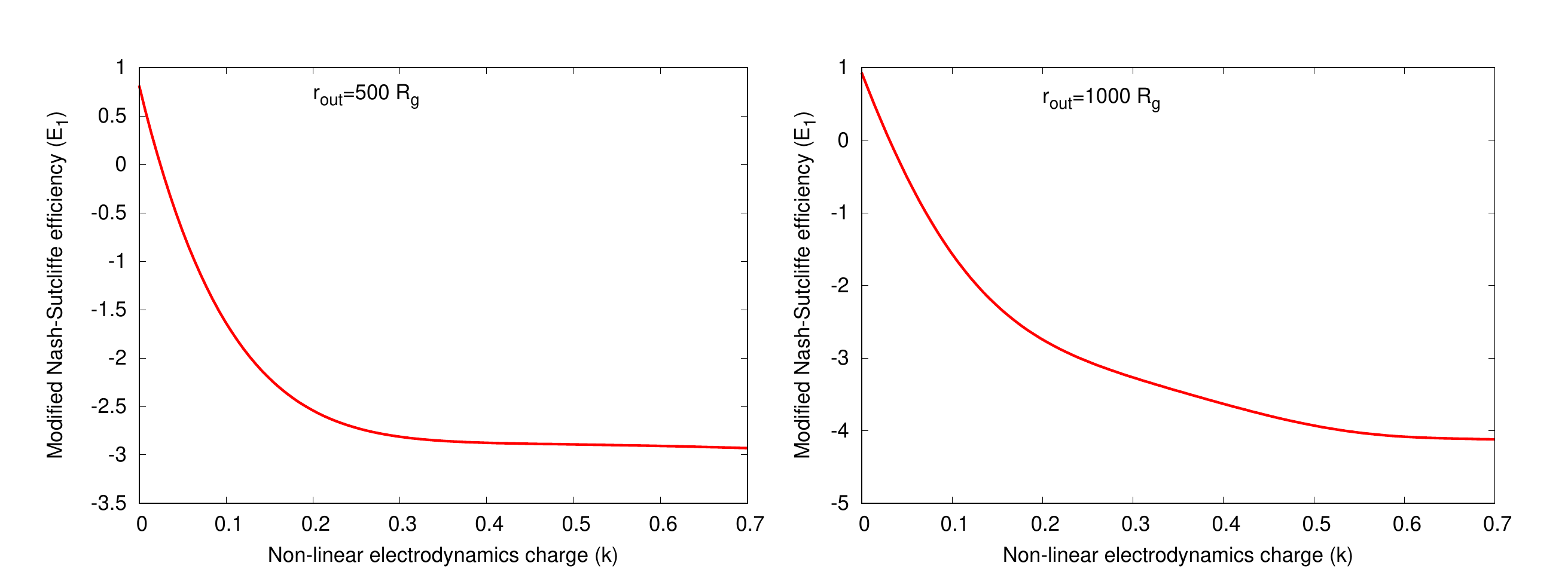}
\end{center}
\caption{Figure 4: In the figure the modified Nash-Sutcliffe Efficiency $E_1$ is plotted with the non-linear electrodynamics charge parameter $k$. In the left panel $E_1$ is plotted for $r_{out}=500r_g$ while $E_1$ calculated with $r_{out}=1000r_g$ is shown in the right panel. In both cases, $E_1$ maximizes for $k=0$ implying that \gr\ is more favored compared to regular black holes in non-linear electrodynamics.}
\label{Fig4}
\end{figure}
From \ref{4-3} it is clear that the value of $k$ where $E_1$ maximizes corresponds to the observationally favored magnitude of $k$. We note from \ref{Fig4} that $k=0$ maximizes $E_1$ which implies that the Kerr scanario is more favored compared to black holes in non-linear electrodynamics. This result holds good irrespective of the choice of $r_{out}$ and is consistent with our previous findings.

\item \textbf{Index of agreement and its modified form:} 
\begin{figure}
\begin{center}
\hspace{-2.1cm}
\includegraphics[scale=0.67]{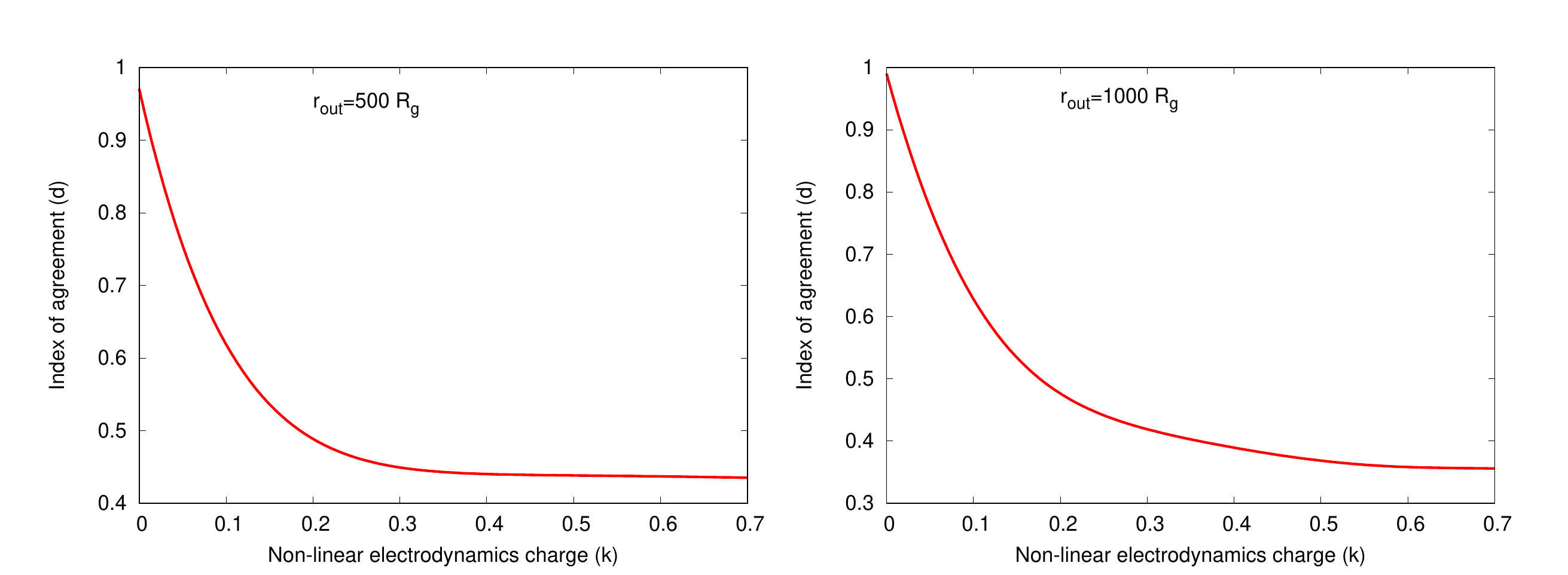}
\end{center}
\caption{Figure 5: The above figure plots the index agreement $d$ with respect to the non-linear electrodynamics charge parameter $k$. The index of agreement computed with $r_{out}=500r_g$ and $r_{out}=1000r_g$ are shown in the left and the right panels respectively. Irrespective of the choice of $r_{out}$, $d$ peaks at $k=0$ supporting the Kerr scenario compared to the regular black hole case.
}
\label{Fig5}
\end{figure}
The Nash-Sutcliffe efficiency and its modified form is insensitive to the differences associated with the theoretical and the observed luminosities from the respective observed mean \cite{WRCR:WRCR8013}. This inspires one to propose two more error estimators, namely the index of agreement and its modified form which takes care of this issue \cite{willmott1984evaluation, doi:10.1080/02723646.1981.10642213,2005AdG.....5...89K}. The index of agreement is given by,
\begin{align}
\label{4-4}
&d(k)=1-\frac{\sum_{j}\left\{{O}_{j}-T_{j}\bigg(k, \lbrace a_{min,j}, \cos i_{min,j},\dot{M}_{min,j} \rbrace \bigg)\right\}^{2}}{\sum _{j}\left\{ \big|{O}_{j}-{O}_{\rm av}\big|+\big|T_{j}\bigg(k,\left \lbrace a_{min,j}, \cos i_{min,j},\dot{M}_{min,j}  \right\rbrace \bigg)-{O}_{\rm av}\big|\right\}^{2}}
\end{align}
In \ref{4-4} the average value of the observed luminosities is given by ${O}_{\rm av}$. The denominator of \ref{4-4} often referred to as the potential error is associated with the maximum departure of each pair of observed and theoretical luminosities from the observed mean ${O}_{\rm av}$. 

Once again, the presence of squared terms in the numerator makes the index of agreement oversensitive to higher values of optical luminosity. As a consequence, a modified version of the index of agreement is proposed which assumes the form,
\begin{align}
\label{4-5}
&d_{1}(k)=1-\frac{\sum_{j}\big|{O}_{j}-T_{j}\bigg(k,\left \lbrace a_{min,j}, \cos i_{min,j},\dot{M}_{min,j} \right\rbrace \bigg)\big|}{\sum_{j}\left\{\big|{O}_{j}-{O}_{\rm av}\big|+\big|T_{j}\bigg(k,\left \lbrace a_{min,j}, \cos i_{min,j},\dot{M}_{min,j}\right\rbrace \bigg)-{O}_{\rm av}\big| \right\}}
\end{align}
\begin{figure}
\begin{center}
\hspace{-2.1cm}
\includegraphics[scale=0.67]{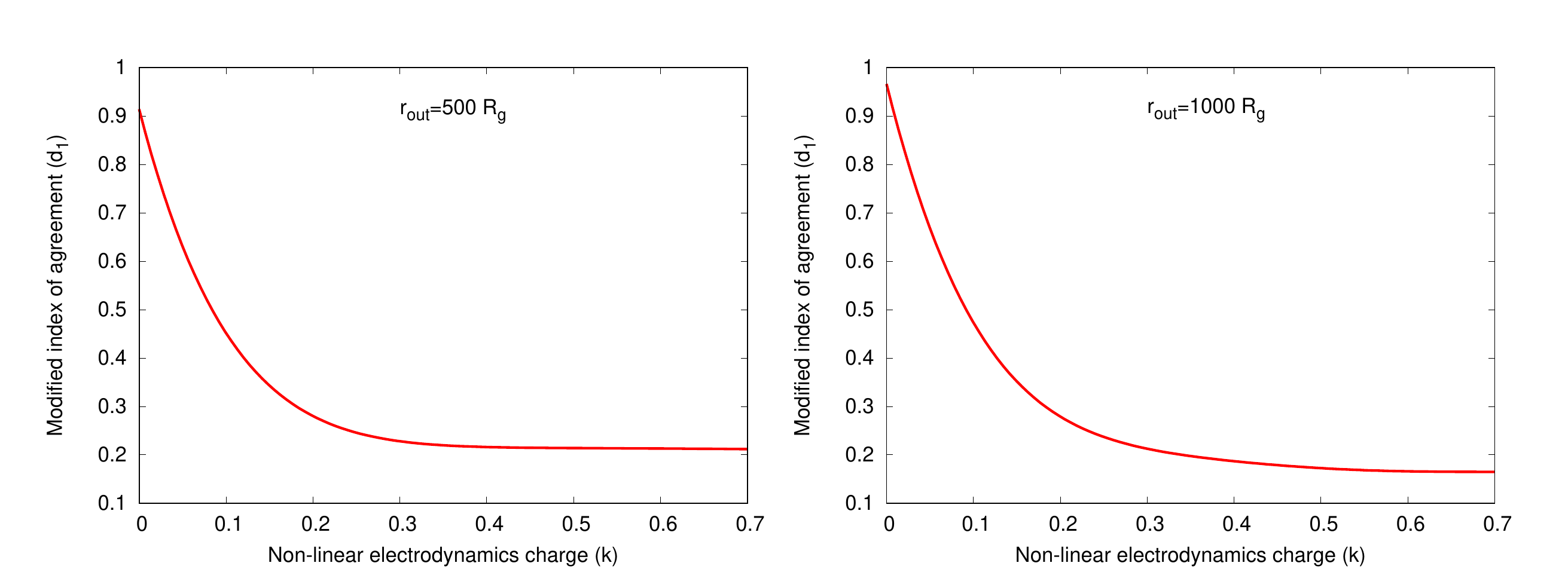}
\end{center}
\caption{Figure 6: The figure above depicts the variation of the modified index of agreement $d_1$ with the non-linear electrodynamics charge parameter $k$. In the left panel $d_1$ is computed with $r_{out}=500r_g$ while in the right panel $d_1$ is evaluated with $r_{out}=1000r_g$. In both cases $d_1$ maximizes for $k=0$ indicating that optical observations of quasars favor the Kerr scenario in \gr\ compared to regular black holes in non-linear electrodynamics.} 
\label{Fig6}
\end{figure}

We note from \ref{4-4} and \ref{4-5} that the magnitude of $k$ where the index of agreement and its modified form maximizes corresponds to the observationally favored value of $k$. \ref{Fig5} and \ref{Fig6} respectively illustrates the variation of $d$ and $d_1$ with respect to $k$ for $r_{out}=500r_g$ (left panel) and $r_{out}=1000r_g$ (right panel). It is clear from the aforesaid figures that $d$ and $d_1$ maximizes when $k=0$ regardless of the choice of $r_{out}$. This is in accordance with our earlier results which states that the Kerr scenario is preferred compared to regular black holes with a Minkowski core as far as optical observations of quasars are concerned. 
Thus, all the five error estimators indicate towards the same conclusion. Interestingly, in an earlier work we studied the signatures of Bardeen blackholes from the present quasar data. Bardeen black holes are examples of regular black holes with a de Sitter core arising in Einstein gravity coupled to non-linear electrodynamics. As in the present case, error analysis showed that quasar optical data favors the Kerr scanario compared to black holes with a non-linear electrodynamics charge \cite{Banerjee:2022chn}. 
In the next section we report the spins of the quasars corresponding to the observationally favored magnitude of $k$ (i.e. $k=0$).

\end{itemize}

\subsection{Spins of the quasars}\label{S4-2}
The present section gives an estimate of the spin of the quasars derived from their optical observations. We recall here that for a given $k$, every quasar has an observationally favored value of spin denoted by $a_{min}$ discussed in \ref{S4}. From the discussion in the last section we note that the $\chi^2$ minimizes while the other errors maximize when $k=0$. We therefore report spins ($a_{min}$) corresponding to $k=0$ for the quasars in \ref{Table1}. It is important to note that the theoretical luminosity depends both on the inner disk radius (which in the present case is the marginally stable circular orbit $r_{ms}$) and the radius of the outer disk $r_{out}$. Since the flux emitted from the accretion disk peaks close to the $r_{ms}$, the luminosity is much more sensitive to the inner radius. Hence, the choice of $r_{out}$ does not affect our observationally favored magnitude of $k$. However, for some quasars the choice of $r_{out}$ does affect the value of $a_{min}$ corresponding to $k=0$. We report spins of only those quasars in \ref{Table1} whose $a_{min}$ corresponding to $k=0$ remains invariant with $r_{out}$. The variation of $a_{min}$ (for a given $k$) with $r_{out}$ stems from the fact that we allowed the inclination angle and the accretion rate to vary while calculating the error estimators. Moreover, the theoretical luminosity is directly related to the temperature profile $T(r)$ which in turn depends on $\dot{M}/\mathcal{M}^2$. Hence the quasars for which this ratio is high, gets some contribution to the theoretical luminosity from the outer disk. This in turn alters the value of $a_{min}$ for those quasars for a given $k$. In \ref{Table1} we report the Kerr parameter of only those quasars whose spin remains unaltered with variation of $r_{out}$.

The spins of some of the quasars presented in \ref{Table1} have been determined independently by other methods. We now compare our spin constraints with that of the previously estimated spins of the quasars. The spin of PG 0003+199 have been determined previously by several authors. For example, according to Keek et al. \cite{Keek:2015apa}
the spin of PG 0003+199 turns out to be $a \sim 0.89\pm 0.05$, while Walton et al. \cite{Walton:2012aw} reported the spin of PG 0003+199 to be $a\sim 0.83^{+0.09}_{-0.13}$.
Based on the general relativistic disk reflection model \cite{Ross:2005dm,Crummy:2005nj}reported that the quasars PG 0003+199, PG 0050+124, PG 1244+026, PG 1404+226, PG 1440+356 are maximally spinning. These results are nearly consistent with our findings. Using polarimetric observations of AGNs \cite{Afanasiev:2018dyv} the spins of PG 0003+199, PG 0050+124, PG 0923+129, PG 2308+098, PG 1022+519, PG 1425+267, PG 1545+210, PG 1613+658 and PG 1704+608 have been determined. We note that our estimate of spin for the quasars PG 0003+199, PG 0050+124, PG 0923+129 and PG 2308+098 are in agreement with their results while our estimate of spin for PG 1022+519, PG 1425+267, PG 1545+210, PG 1613+658 and PG 1704+608 show some deviations. Bottacini et al. \cite{Bottacini:2014lva} reported that PG 1613+658 (Mrk 876) consists of a rotating black hole which is in accordance with our findings. According to \cite{2017Ap&SS.362..231P} the spin of PG 1704+608 (3C 351) is $a<0.998$ which is in agreement with our results.

\begin{table}

\caption{Table 1: Spin parameters of quasars corresponding to $k=0$}
\label{Table1}
\vspace{-0.36cm}
\begin{center}
\centering
\begin{tabular}{|c|c|c|}

\hline
$\rm Object$ & $\rm log~ m$ & $a_{k=0}$\\
\hline
$\rm 0003+199$ & $\rm 6.88$  &  $\rm 0.99 $\\ \hline
$\rm 0050+124$ & $\rm 6.99$  &  $\rm 0.99 $\\ \hline
$\rm 0923+129$ & $\rm 6.82$  &  $\rm 0.99 $\\ \hline
$\rm 1011-040$ & $\rm 6.89$ &   $\rm 0.99 $\\ \hline
$\rm 1022+519$ & $\rm 6.63$ &   $\rm 0.99$\\ \hline
$\rm 1119+120$ & $\rm 7.04$ &   $\rm 0.99 $\\ \hline
$\rm 1244+026$ & $\rm 6.15$ &   $\rm 0.99 $\\ \hline
$\rm 1404+226$ & $\rm 6.52$ &   $\rm 0.99 $\\ \hline
$\rm 1425+267$ & $\rm 9.53$ &   $\rm 0.7 $\\ \hline
$\rm 1440+356$ & $\rm 7.09$ &   $\rm 0.99 $\\ \hline
$\rm 1535+547$ & $\rm 6.78$ &   $\rm 0.99 $\\ \hline
$\rm 1545+210$ & $\rm 9.10$ &   $\rm 0.1 $\\ \hline
$\rm 1552+085$ & $\rm 7.17$  &  $\rm 0.99 $\\ \hline
$\rm 1613+658$ & $\rm 8.89$  &  $\rm 0.3 $\\ \hline
$\rm 1704+608$ & $\rm 9.29$ &   $\rm -0.4 $\\ \hline
$\rm 2308+098$ & $\rm 9.43$ &  $\rm 0.95 $\\ \hline
\end{tabular}
\end{center}
\end{table}

\par

\section{Concluding Remarks}\label{S5}
The present paper investigates the nature of the continuum spectrum emitted by the accretion disk around regular black holes and aims to discern the viability of the regular black hole scenario compared to the Kerr scenario in GR. The regular black holes considered here are endowed with a Minkowski core and arises in Einstein gravity coupled to non-linear electrodynamics. Exploring the signatures of such black holes in astrophysical observations is interesting because they can potentially evade the $r=0$ curvature singularity arising in GR. 
The regular black hole considered here has an exponential mass function and is characterized by the non-linear electrodynamics charge parameter $k$ and the spin parameter $a$. The continuum spectrum depends both on the metric parameters as well as the properties of the accretion flow. We work out the theoretical luminosity in such a background using the thin-disk approximation. This when compared with the optical observations of eighty Palomar Green quasars reveal that the Kerr scenario is observationally much more preferred compared to the present regular black hole scenario. In particular, $\chi^2$ rises very sharply for a non-zero $k$ which indicates that $k>0$ is ruled out outside 3-$\sigma$. It is important to note that the Kerr solution arises in some alternative gravity models as well \cite{Psaltis:2007cw}, and observational confirmation of the Kerr metric may not always indicate that GR is more favored.
In an earlier work we performed the same analysis with Bardeen black holes (regular black holes with a de Sitter core arising in non-linear electrodynamics) \cite{Banerjee:2022chn} and arrived at a similar conclusion. Our findings are based on error estimators like $\chi^2$, the Nash-Sutcliffe efficiency, the index of agreement and their modified forms. Thus black holes in non-linear electrodynamics seem to be disfavored by optical observations of quasars.
The present analysis also allows us to establish constrains on the spins of some quasars which are more or less in agreement with previous estimates. 

Before we conclude we would like to mention some of the shortcomings of the present analysis. First of all, the spectral energy distribution (SED) of quasars consists of contribution from the accretion disk, the corona, the jet and the dusty torus which are difficult to observe and model \cite{Brenneman:2013oba}. Deciphering the signatures of the various components from the SED is not easy which turns out to be a limitation in precise determination of the black hole parameters like mass, distance, spin or inclination. Second, the continuum spectrum depends not only on the background metric but also on the characteristics of the accretion flow. The thin-disk approximation assumes motion along the equatorial plane with negligible radial and vertical velocity. In order to model a realistic accretion flow these assumptions need to be relaxed. This requires a much more detailed modelling of the spectrum taking into account the dynamical evolution of the disk, the corona, the jet and the outflows. At present these issues are addressed by several phenomenological models which is beyond the scope of this work. 
Moreover, the present result is dependent on the observational sample. It is therefore instructive to carry out the present analysis with a different black hole sample.
We have studied the prospect of the present regular black hole scenario with observations related to quasiperiodic oscillations. Such a study reveals that most QPO models favor a vanishing or small value of $k$. Apart from using a different black hole sample, it is worthwhile to investigate the viability of the regular black hole scenario with other astrophysical observations, namely the black hole shadow, Fe-line method and so on. This in turn can establish independent constraints on the non-linear electrodynamics charge parameter, which can be compared with the present work. We leave this for a future work which will be reported elsewhere.

.


\bibliography{accretion,accretion2,torsion,Gravity_1_full,Gravity_2_full,Gravity_3_partial,Brane,KN-ED,KN-ED2,EMDA-Jet,bardeen,Black_Hole_Shadow,regularBh,new-ref,ref,QG,IB,QPO}

\bibliographystyle{./utphys1}

\end{document}